\documentclass[journal,twocolumn,10pt,twoside]{IEEEtran}

\usepackage[T1]{fontenc}
\usepackage[latin9]{inputenc}
\usepackage{amsthm} 

\usepackage{graphicx}
\usepackage{setspace,bm}
\usepackage{bbm}
\usepackage{framed,cite}
\usepackage{psfrag}
\usepackage{multicol}
\usepackage{multirow}
\usepackage{blindtext}
\usepackage{amsmath,amssymb} 
\usepackage{color}
\usepackage{tikz}
\usetikzlibrary{decorations.pathreplacing}
\usepackage{pgfplots} 
\pgfplotsset{width=7cm}
\usepackage[caption=false,font=footnotesize,subrefformat=parens,labelformat=parens]{subfig}
\usepackage[normalem]{ulem}
\usepackage{algorithm}
\usepackage{algpseudocode}
\usepackage{enumitem} 

\usetikzlibrary{shapes.multipart,matrix,positioning}

\newcommand{\vecx}{\mathbf{x}}
\newcommand{\vecy}{\mathbf{y}}

\newcommand{\vecz}{\mathbf{z}}
\newcommand{\vecs}{\mathbf{s}}
\newcommand{\vecX}{\mathbf{X}}
\newcommand{\vecY}{\mathbf{Y}}
\newcommand{\vecZ}{\mathbf{Z}}

\newcommand{\varNumPart}{N_p}
\newcommand{\lenX}{K} 
\newcommand{\lenY}{J} 
\newcommand{\spanNum}{N} 
\newcommand{\lenMC}{N_\textrm{mc}} 
\newcommand{\alpX}{\mathcal X} 
\newcommand{\alpY}{\mathcal Y} 

\newcommand{\jlt}{J. Lightw. Technol.}
\newcommand{\ptl}{IEEE Photon. Technol. Lett.}
\newcommand{\ope}{Optics Express}
\newcommand{\tit}{IEEE Trans. Inf. Theory}
\newcommand{\tcom}{IEEE Trans. Commun.}
\newcommand{\ofc}{Proc. Optical Fiber Communication Conference (OFC)}

\newcommand{\ecoc}{Proc. European Conference on Optical Communication (ECOC)}

\usepackage{amsthm}
\newtheoremstyle{break}{20pt}{5pt}{\itshape}{}{}{:}{\newline}{}
\theoremstyle{break}

\normalsize

\ifCLASSINFOpdf
\else
\fi

\begin{document}
\title{{Improved Lower Bounds on Mutual Information Accounting for Nonlinear Signal--Noise Interaction}} 


\author{Naga V. Irukulapati, Marco Secondini, \IEEEmembership{Senior Member, IEEE}, Erik Agrell, \IEEEmembership{Fellow, IEEE}, Pontus Johannisson, and Henk Wymeersch, \IEEEmembership{Member, IEEE}
\thanks{{N.~V.~Irukulapati was with the Dept. of Electrical Engineering at Chalmers University of Technology, Gothenburg, Sweden and is now working at Ericsson Research, Ericsson, Gothenburg, Sweden. H. Wymeersch and E. Agrell are with the Dept. of Electrical Engineering at Chalmers University of Technology, Gothenburg Sweden. 
M. Secondini is with the Institute of Communication, Information, and Perception Technologies, Scuola Superiore Sant'Anna, Pisa, Italy. P. Johannisson is with RISE Acreo Gothenburg, Sweden. Authors email: irukulapati.nagav@gmail.com, marco.secondini@sssup.it, agrell@chalmers.se, pontus.johannisson@ri.se, henkw@chalmers.se}}
\thanks{This research was supported by the Swedish Research Council (VR) under grant 2013-5642 and the European Research Council under grant no. 258418 (COOPNET). The simulations were performed in part on resources provided by the Swedish National Infrastructure for Computing (SNIC) at C3SE.}
}

\maketitle

\begin{abstract}
In fiber-optic communications, evaluation of mutual information (MI) is still an open issue due to the unavailability of an exact and mathematically tractable channel model. Traditionally, lower bounds on MI are computed by approximating the (original) channel with an auxiliary forward channel. In this paper, lower bounds are computed using an auxiliary backward channel, which has not been previously considered in the context of fiber-optic communications. Distributions obtained through two variations of the stochastic digital backpropagation (SDBP) algorithm are used as auxiliary backward channels and these bounds are compared with bounds obtained through the conventional digital backpropagation (DBP). Through simulations, higher information rates were achieved with SDBP, {which can be explained by the ability of SDBP to account for nonlinear signal--noise interactions}.
\end{abstract}


\IEEEpeerreviewmaketitle

\section{Introduction}\label{secIntro}

Shannon proved that reliable communication through a noisy channel is possible with channel coding, as long as the information rate is less than {the} channel capacity \cite{Shannon1948}. 
For any fixed input distribution, {the} mutual information (MI) gives a lower bound on the channel capacity. 
MI is also shown to be a better metric than the {pre-forward-error-correction} {bit-error rate} for estimating the {post-forward-error-correction} {bit-error rate} in soft-decision {forward-error-correction} systems \cite{Wan2006,Franceschini2006,Brueninghaus2005,Leven2011}. 
For a discrete-time channel with memory, the MI between random vectors $\vecX \triangleq (X_1, X_2,\ldots,X_\lenX)$ and $\vecY \triangleq (Y_1, Y_2,\ldots,Y_\lenY)$ with $\lenY \ge \lenX$ is defined as\footnote{{All logarithms in this paper are in base $2$; therefore, MI will be measured in bits. To simplify the notation, we used $p(\vecy|\vecx)$, $p(\vecx)$, and $p(\vecy)$ instead of explicitly writing $p_{\vecY|\vecX}(\vecy|\vecx)$, $p_{\vecX}(\vecx)$, and $p_{\vecY}(\vecy)$. So, $p(\vecy|\vecx)$, $p(\vecx)$, $p(\vecy)$, and their auxiliary counterparts refer to different distributions. For the limit in (\ref{eqnMIrate}) to exist, we assume the existence of sequences of distributions $p(\vecy|\vecx)$, $p(\vecx)$, and $p(\vecy)$ for $\lenX=1,2,\ldots$}}
\begin{align}
\label{eqnMIXnYnpyx} \hspace{-0.15cm} I(\vecX; \vecY)= \mathbb E_{\vecX,\vecY}\bigg[\log\frac{p(\vecY|\vecX)}{p(\vecY)}\bigg]= \mathbb E_{\vecX,\vecY}\bigg[\log\frac{p(\vecX|\vecY)}{p(\vecX)}\bigg]
\end{align}
where $\vecx = (x_1,x_2,\ldots,x_\lenX) \in \alpX$ is a realization of $\vecX$ drawn from the input distribution $p(\vecx)$, and $\vecy = (y_1,y_2,\ldots,y_\lenY) \in \alpY$ is a realization of the corresponding output random vector $\vecY$. $p(\vecy|\vecx)$ is the channel conditional distribution,  $p(\vecy)=\int_{\alpX} p(\vecx)p(\vecy|\vecx)\mathrm{d}\vecx$ is the output distribution, and $\mathbb E_{\vecX,\vecY}[.]$ is expectation over the joint distribution $p(\vecx,\vecy)=p(\vecy|\vecx)p(\vecx)=p(\vecx|\vecy)p(\vecy)$. 
The information rate between the ergodic processes for the channel with memory is \cite{Arnold2006}
\begin{align}\label{eqnMIrate}
I^{\textrm{mem}}  = \lim_{\lenX,\lenY \rightarrow \infty} \frac{1}{\lenX}I(\vecX; \vecY).
\end{align}

It is often the case, especially for the fiber-optic channel, that the channel distributions, $p(\vecy|\vecx)$ and $p(\vecx|\vecy)$, are not known in closed form. Hence, the MI of (\ref{eqnMIXnYnpyx}) and subsequently the information rate of (\ref{eqnMIrate}) cannot be computed in closed form.  The information rate of (\ref{eqnMIrate}) can, in principle, be estimated through simulations using the forward recursion of the BCJR algorithm \cite{Arnold2006}. {The complexity of this simulation-based technique increases exponentially with channel memory and  requires knowledge of the channel model. There are at least three different ways in which the problem of the exponential memory for the BCJR algorithm is currently addressed: limiting the memory by truncation \cite{Colavolpe2011, Lin2015}; using digital backpropagation (DBP) before the BCJR algorithm \cite{Djordjevic2010, Sugihara2013}; or using low-complexity variations of the BCJR algorithm to reduce the complexity. By employing one of these techniques, the information rate obtained will be a lower bound on the maximum achievable information rate. The problem of not knowing the exact channel model is solved by using an approximation of the channel model \cite{Djordjevic2005, Colavolpe2011, Lin2015, Secondini2013,Essiambre2010Capacity}. {In this paper, instead of using the forward recursion of the BCJR algorithm, we use distributions obtained from stochastic digital backpropagation (SDBP) \cite{Irukulapati2014TCOM}, which accounts for the memory of the fiber-optic channel channel and provides an approximate posterior distribution that can be used to compute lower bounds on the MI. In contrast to the forward recursion of the BCJR algorithm, the complexity of SDBP
does not grow exponentially as a function of the channel memory.} } {
In \cite{Essiambre2010Capacity}, ring constellations are used as input distributions and a memoryless channel is assumed after DBP is applied to the center channel in a wavelength division multiplexing system. 
In \cite{Secondini2013},  a new channel model for a wavelength division multiplexed system (after using DBP for the channel of interest) is introduced and an AIR is computed using this channel model as an auxiliary forward channel. In \cite{Agrell2014}, a channel based on a finite-memory Gaussian noise model is studied, which is different from the channel used in this paper.}

{Most of these techniques fall into the category of mismatched decoding \cite{Fischer1978,Merhav1994}.} In mismatched decoding, the original distributions $p(\vecy|\vecx)$ or $p(\vecx|\vecy)$ are approximated with auxiliary distributions and the rates computed using these auxiliary distributions are a lower bound on the MI. The {better} the auxiliary distribution approximates the original distribution, the {closer} are these bounds to the MI. Before proceeding, we define four entities, similar to \cite[Fig.~1]{Sadeghi2009}, that will be used throughout the paper:
\begin{itemize}
    \item $p(\vecy|\vecx)$ is the original forward channel;
    \item $p(\vecx|\vecy)$ is the original backward channel;
    \item $q(\vecy|\vecx)$ is an  auxiliary forward channel; 
    \item $r(\vecx|\vecy)$ is an auxiliary backward channel; 
\end{itemize}
We define a backward channel by reversing the usual meaning of $\vecX$ and $\vecY$, i.e., looking at $\vecX$ as being the output of some channel which is fed by $\vecY$, which in turn is produced by some source \cite{Merhav1993,Sadeghi2009}. 
Note that the original backward channel is associated with the original forward channel {using Bayes' rule} as $p(\vecx|\vecy)=p(\vecy|\vecx)p(\vecx)/p(\vecy)$. However, it is not necessary that such a relation exists between auxiliary channels, i.e., $r(\vecx|\vecy)$ can be any conditional distribution which does not correspond\footnote{To highlight this difference, we chose to use $r(\vecx|\vecy)$ for an auxiliary backward channel instead of $q(\vecx|\vecy)$.} to any auxiliary forward channel $q(\vecy|\vecx)$. {We discuss this problem in more detail in Sec. \ref{secRevAuxCh}}. 

{For the fiber-optic channel, }the most commonly used approach to lower-bound the MI is to approximate the original forward channel with an auxiliary forward channel {\cite{Djordjevic2005,Colavolpe2011,Lin2015,Leven2011,Secondini2013,Fehenberger2015a,Eriksson2016,Liga2016}}. A receiver that is optimal for an auxiliary forward channel is used to process the data generated from the original forward channel {and to compute an information rate. This rate is achievable by that receiver and, for this reason, is often referred to as an achievable information rate (AIR)} \cite{Ganti2000,Arnold2006,Secondini2013}.  
{In \cite{Fehenberger2015a, Liga2016}, a circularly symmetric AWGN channel is used as an auxiliary channel and AIRs are computed both for soft-decision and hard-decision systems. In \cite{Eriksson2016}, AIRs are computed using 4-dimensional Gaussian distributions as auxiliary channel. Both techniques are evaluated in this paper for benchmarking purposes. }

An alternative approach to lower-bound the MI is through a direct use of the auxiliary backward channel. {The concept of auxiliary backward channel was used for the first time in the context of universal decoding for memoryless channels with deterministic interference \cite{Merhav1993}.} An auxiliary backward channel is used instead of an auxiliary forward channel to maximize the lower bound using an iterative procedure \cite{Sadeghi2009}.

{\textit{Contributions of the paper}: In this paper, lower bounds are computed using an auxiliary backward channel, which has not been previously considered in the context of fiber-optic communications.} 
{Specifically, distributions obtained from two variations of the SDBP algorithm are used as auxiliary backward channels, namely from symbol-by-symbol SDBP (SBS-SDBP) \cite{Irukulapati2014TCOM} and Gaussian message passing SDBP (GMP-SDBP) \cite{Wymeersch2015SPAWC}. 
Through simulations, the AIR computed using these two variants of SDBP was observed to be higher than the AIR obtained using the conventional DBP algorithm.} 

\textit{Organization of the paper}: {Similarities and differences in the computation of lower bounds on the MI using an auxiliary forward channel and auxiliary backward channel are described in Sec.~\ref{secMILowerBounds}}. Computation of AIRs for the fiber-optic channel is considered in Sec.~\ref{secAuxSDBP}, where SBS-SDBP and GMP-SDBP are described briefly, leading to a discussion on how auxiliary backward channels are obtained using these two approaches. AIRs computed using these two versions of SDBP are then compared with DBP. Numerical results are presented and discussed in Sec.~\ref{secNumSim}, followed by conclusions in Sec.~\ref{secConcl}.

\section{Lower bounds on mutual information}\label{secMILowerBounds}


\subsection{Lower Bounds using Auxiliary Forward Channel $q(\vecy|\vecx)$}\label{subSecLBauxiliary forward channel}
{A lower bound on the MI using an auxiliary forward channel is} \cite[Eq.~(41)]{Arnold2006}
\begin{align}\label{eqnIqyxleqIxy}
I(\vecX;\vecY)  \geq I_q(\vecX;\vecY) =  \mathbb E_{\vecX,\vecY}\bigg[\log\frac{q(\vecY|\vecX)}{q(\vecY)}\bigg] 
\end{align}
where $q(\vecy)\triangleq \int_{\alpX} p(\vecx)q(\vecy|\vecx)\mathrm{d}\vecx$ is the output distribution obtained by connecting the original source $p(\vecx)$ to the auxiliary forward channel. {This lower bound, $I_q(\vecX;\vecY)$ of (\ref{eqnIqyxleqIxy}), can be achieved by using a maximum a posteriori detector designed for the auxiliary forward channel and used as a receiver for the original forward channel \cite{Fischer1978,Ganti2000}. {Since the data generated by the original forward channel is processed by a receiver that is optimized for a different auxiliary forward channel, this approach is known as mismatched decoding.}}

Using the {Kullback--Leibler} divergence \cite[Sec.~8.1]{Cover2006}, it can be easily verified {\cite[Eq.~(34)--(41)]{Arnold2006}} that (\ref{eqnIqyxleqIxy}) is indeed a lower bound,
\begin{align}\label{eqnKLDgeq0}
I(\vecX;\vecY) - I_q(\vecX;\vecY)=D(p(\vecx,\vecy)||p(\vecy)r_q(\vecx|\vecy)) \geq 0,
\end{align}
where
\begin{align}\label{eqnrqbyauxiliary forward channel}
r_q(\vecx|\vecy) \triangleq \frac{p(\vecx)q(\vecy|\vecx)}{q(\vecy)}
\end{align}
is the auxiliary backward channel induced by the auxiliary forward channel $q(\vecy|\vecx)$. A sufficient condition for the inequality (\ref{eqnKLDgeq0}) to hold is that $p(\vecy) r_q(\vecx|\vecy)$ is a joint distribution, i.e., $\int_\alpX \int_\alpY p(\vecy) r_q(\vecx|\vecy) \mathrm d\vecy \mathrm d\vecx=1$ \cite[Th. 8.6.1]{Cover2006}. {This condition is fulfilled for any combination of $p(\vecy|\vecx)$ and $q(\vecy|\vecx)$, and can be verified by using $q(\vecy)= \int_{\alpX} p(\vecx)q(\vecy|\vecx)\mathrm{d}\vecx$ in $r_q(\vecx|\vecy)$ of (\ref{eqnrqbyauxiliary forward channel}).}

\subsection{Lower Bounds using Auxiliary Backward Channel $r(\vecx|\vecy)$}\label{secRevAuxCh}
 
There are instances such as SDBP, where an auxiliary backward channel $r(\vecx|\vecy)$ is known while the corresponding auxiliary forward channel $q(\vecy|\vecx)$ is unknown. In such cases, if (\ref{eqnIqyxleqIxy}) is to be used to compute a lower bound on the MI, an auxiliary forward channel $q(\vecy|\vecx)$ has to be computed corresponding to a given auxiliary backward channel $r(\vecx|\vecy)$ and a given $p(\vecx)$. There are two challenges with this approach. Firstly, given $r(\vecx|\vecy)$ and $p(\vecx)$, no general method exists to compute $q(\vecy|\vecx)$, and depending on the input and output alphabets, there may not always exist a corresponding $q(\vecy|\vecx)$ or there may exist multiple solutions. {We show that given an auxiliary backward channel, an auxiliary forward channel may not always exist, by providing a counter-example. Given $r(\mathbf{x}|\mathbf{y})$ and $p(\mathbf{x})$, the problem is now to find $q(\mathbf{y}|\mathbf{x})$ for all $\mathbf{y}$ and $\mathbf{x}$.  Suppose we have a detector that provides  $r(\mathbf{x}|\mathbf{y})=r(\mathbf{x})\neq p(\mathbf{x})$, irrespective of $\mathbf{y}$. For such a detector, the auxiliary forward channel induced by $r(\mathbf{x}|\mathbf{y})$ is given by $q(\mathbf{y}|\mathbf{x})=r(\mathbf{x}|\mathbf{y})q(\mathbf{y})/p(\mathbf{x})=r(\mathbf{x})q(\mathbf{y})/p(\mathbf{x})$. The condition $\int q(\mathbf{y}|\mathbf{x}) \mathrm{d}\mathbf{y}=1$ then implies $r(\mathbf{x})=p(\mathbf{x})$, which is contradictory to the assumption $r(\mathbf{x})\neq p(\mathbf{x})$. Hence, not every auxiliary backward channel has an associated auxiliary forward channel. Secondly, even when there exists a $q(\mathbf{y}|\mathbf{x})$ corresponding to a given auxiliary backward channel and $p(\vecx)$, it may be computationally intractable to obtain from $r(\mathbf{x}|\mathbf{y})$ for all $\mathbf{y}$, since $\mathbf{y}$ can take on uncountably many values, while $r(\mathbf{x}|\mathbf{y})$ is only available for specific observed values of $\mathbf{y}$. } In this section, we provide an alternate approach to lower-bounding the MI using an auxiliary backward channel, without the explicit knowledge of a corresponding auxiliary forward channel. 

For any input distribution $p(\vecx)$ and any conditional distribution $r(\vecx|\vecy)$, lower bounds on the MI can be derived as \cite[Eq.~(39)]{Sadeghi2009}
\begin{align}\label{eqnIqxyleqIxy}
I(\vecX;\vecY) \geq I_r(\vecX;\vecY) = \mathbb E_{\vecX,\vecY}\bigg[\log\frac{r(\vecX|\vecY)}{p(\vecX)}\bigg] ,
\end{align}
where, similarly to (\ref{eqnIqyxleqIxy}), the averaging is computed with respect to the joint distribution $p(\vecx,\vecy)$. The lower bound (\ref{eqnIqxyleqIxy}) can be also proved using {Kullback--Leibler} divergence as
\begin{align}\label{eqnKLDRrgeq0}
I(\vecX;\vecY)- I_r(\vecX;\vecY)=D(p(\vecx,\vecy)||{p(\vecy)r(\vecx|\vecy)}) \geq 0 
\end{align}
For the inequality in (\ref{eqnKLDRrgeq0}) to hold, $p(\vecy)r(\vecx|\vecy)$ should be a joint distribution, which is always true as long as $r(\vecx|\vecy)$ is chosen as a conditional distribution. Note that (\ref{eqnIqyxleqIxy}) and (\ref{eqnIqxyleqIxy}) provide the same lower bound, i.e., $I_q(\vecX;\vecY)=I_r(\vecX;\vecY)$, {only} if the auxiliary backward channel is given by $r_q(\vecx|\vecy)$, i.e., induced by the auxiliary forward channel according to (\ref{eqnrqbyauxiliary forward channel}). However, it should be noted that (\ref{eqnIqxyleqIxy}) can be used as a lower bound on the MI for any arbitrary conditional distribution $r(\vecx|\vecy)$, which is not necessarily related to any $q(\vecy|\vecx)$. 
It has been shown that $I_r(\vecX;\vecY)$ is achievable using an optimal detector, i.e., a maximum a posteriori detector, designed for the auxiliary backward channel $r(\vecx|\vecy)$ \cite{Ganti2000},\cite[Eq.~{(42)--(43)}]{Sadeghi2009}. {The decisions $\hat{\vecx}$ are taken as $\hat{\vecx}=\arg \max_{\vecx} r(\vecx|\vecy)$.}

\textit{Remark 1}: {Either maximizing $I_q$ of (\ref{eqnIqyxleqIxy}) over all possible auxiliary forward channels $q(\vecy|\vecx)$ or maximizing $I_r$ of (\ref{eqnIqxyleqIxy}) over all possible auxiliary backward channels $r(\vecx|\vecy)$ leads to the true MI {(\ref{eqnMIXnYnpyx})}.}

\subsection{Monte Carlo Estimation of AIR}

By combining (\ref{eqnMIrate}) and (\ref{eqnIqyxleqIxy}), and (\ref{eqnMIrate}) and (\ref{eqnIqxyleqIxy}), we have
\begin{align}
\label{eqnIrateyx} I^{\textrm{mem}}_q = \lim_{\lenX,\lenY \rightarrow \infty} \frac{1}{\lenX} \mathbb E_{\vecX,\vecY}\bigg[\log\frac{q(\vecY|\vecX)}{q(\vecY)}\bigg], \\
\label{eqnIratexy} I^{\textrm{mem}}_r = \lim_{\lenX,\lenY \rightarrow \infty}\frac{1}{\lenX} \mathbb E_{\vecX,\vecY}\bigg[\log\frac{r(\vecX|\vecY)}{p(\vecX)}\bigg].
\end{align}

The state-of-the-art method for the estimation of (\ref{eqnIrateyx}) and (\ref{eqnIratexy})  is a simulation based on Monte Carlo (MC) averages \cite{Arnold2001, Arnold2006}. 
The channel is simulated $\lenMC$ times, each time by generating input $\vecx$ with different random seed, say $\vecx^{(n)}$ for the $n$th MC run, to get a corresponding output $\vecy^{(n)}$  from the fiber-optic channel for each MC run. 
The lower bound on the MIs (\ref{eqnIrateyx}) and (\ref{eqnIratexy}) can then be estimated as
\begin{align}
\label{eqnMIMC_qyx} \hat{I}_q^{\textrm{mem}} &= \frac{1}{\lenMC} \sum_{n=1}^{\lenMC}  \left\{\frac{1}{\lenX} \log \frac{q(\vecy^{(n)}|\vecx^{(n)})}{q(\vecy^{(n)})}\right\}, \\
\label{eqnMIMC_Rxy} \hat{I}_r^{\textrm{mem}} &= \frac{1}{\lenMC} \sum_{n=1}^{\lenMC}  \left\{\frac{1}{\lenX} \log \frac{r(\vecx^{(n)}|\vecy^{(n)})}{p(\vecx^{(n)})}\right\}. 
\end{align}

\begin{figure*}
\centering
\begin{tikzpicture}

\fill [fill=lightgray] (-3.6,-1.3) rectangle (6.1,0.15);

\draw[-][blue,thick,dotted] (-3.6,-1.2) -- (-3.6,1.6) node at (-2.5,1.8) {GMP-SDBP};
\draw[-][blue,thick,dotted] (-3.6,1.6) -- (6.1,1.6) ;
\draw[-][blue,thick,dotted] (6.1,0.4) -- (6.1,1.6) ;
\draw[-][blue,thick,dotted] (-0.8,0.4) -- (6.1,0.4) ;
\draw[-][blue,thick,dotted] (-0.8,-1.2) -- (-0.8,0.4) ;
\draw[-][blue,thick,dotted] (-3.6,-1.2) -- (-0.8,-1.2) ;

\fill [fill=blue!20!white] (-0.8,0.4) rectangle (6.1,1.6);
\fill [fill=blue!20!white] (-3.6,-1.2) rectangle (-0.8,1.6);

\draw [black, thick, dashed] (-3.6,-1.3) rectangle (6.1,0.15) node at (-2.5,0.4) {SBS-SDBP};

\draw [black, thick] (-0.25,0.5) rectangle (2,1.5) node at (0.9,1) {GMP};
\draw[-][thick] (-1,1) -- (-0.25,1); 
\draw[-][thick] (2,1) -- (3.75,1) node at (3,1.2) {$\tilde{r}_k(\cdot|\vecy)$}; 
\draw [black, thick] (3.75,0.5) rectangle (6,1.5) node at (4.9,1.32) {Evaluate in $\Omega$,} node at (4.9,0.92) {multiply prior,} node at (4.9,0.65) {normalize};
\draw[->][thick] (6,1) -- (7,1) node at (6.8,1.3) {$r_k(\cdot|\vecy)$};
\draw[-][thick] (-1,-0.5) -- (-1,1); 

\draw [black, thick] (-0.25,0) rectangle (2,-1) node at (0.9,-0.25) {MF+samp.+} node at (0.9,-0.75) {Gauss. approx.};
\draw[-][thick] (-1.5,-0.5) -- (-0.25,-0.5); 
\draw[-][thick] (2,-0.5) -- (3.75,-0.5) node at (3,-0.25) {$\tilde{r}_k(\cdot|\vecy)$}; 
\draw [black, thick] (3.75,0) rectangle (6,-1) node at (4.9,-0.18) {Evaluate in $\Omega$,} node at (4.9,-0.58) {multiply prior,} node at (4.9,-0.85) {normalize}; 
\draw[->][thick] (6,-0.5) -- (7,-0.5) node at (6.8,-0.2) {$r_k(\cdot|\vecy)$};

\draw [black, thick] (-1.5,0) rectangle (-3.5,-1) node at (-2.5,-0.5) {SDBP};
\draw[-][thick] (-5,-0.5) -- (-3.5,-0.5)  node at (-4.25,-0.25) {$\vecy$} node at (-4.25,-0.75) {$\vecy \in \mathbb C^{\lenY}$} ;
\draw [black, thick] (-5,0) rectangle (-7,-1) node at (-6,-0.5) {Fiber link};
\draw[-][thick] (-7.5,-0.5) -- (-7,-0.5) node at (-7.25,-0.25) {$\vecs$} node at (-7.25,-1.25) {$\vecs \in \mathbb C^\lenY$} ;
\draw [black, thick] (-7.5,0) rectangle (-9.5,-1) node at (-8.5,-0.5) {Pulse shaper};
\draw[-][thick] (-11,-0.5) -- (-9.5,-0.5) node at (-10.25,-0.25) {$\vecx$} node at (-10.25,-0.75) {$\vecx \in \Omega^\lenX$} node at (-10.25,-1.25) {$\Omega \subseteq \mathbb C$} ;

\draw[decorate,decoration={brace,amplitude=10pt}] (-9.5,0.25) -- coordinate[below=10pt](C) (-5,0.25) node at (-7.2,0.8) {fiber-optic channel};

\draw [black, thick] (-1.5,-1.5) rectangle (-3.5,-2.5) node at (-2.5,-2) {DBP};
\draw[-][thick] (-4.3,-2) -- (-3.5,-2); 
\draw[-][thick] (-4.3,-2) -- (-4.3,-0.5); 
\draw[-][thick] (-1.5,-2) -- (-0.25,-2) ;
\draw [black, thick] (2,-1.5) rectangle (-0.25,-2.5) node at (1,-2) {MF+samp.};
\draw[-][thick] (2,-2) -- (3.75,-2) node at (2.8,-1.8) {$\vecz$} node at (3,-2.3) {$\vecz \in \mathbb C^{\lenX}$};
\draw [black, thick] (6,-1.5) rectangle (3.75,-2.5) node at (4.9,-1.8) {Estimate pars.} node at (5,-2.2) {for Gauss.};
\draw[->][thick] (6,-2) -- (7,-2) node at (6.8,-1.7) {$q(z_k|x_k)$};

\end{tikzpicture}
\caption{\label{figDiffAuxChanlSDBP} Auxiliary channels obtained for the fiber-optic channel using two variations of SDBP, and DBP, where {$x_k \in \Omega$}. {The distribution $\tilde{r}_k(\cdot|\vecy)$ is evaluated at $x_k \in \Omega$, {multiplied with the prior,} and normalized to get an auxiliary backward channel, ${r}_k(\cdot|\vecy)$. {MF refers to the matched filter}.} }
\end{figure*}
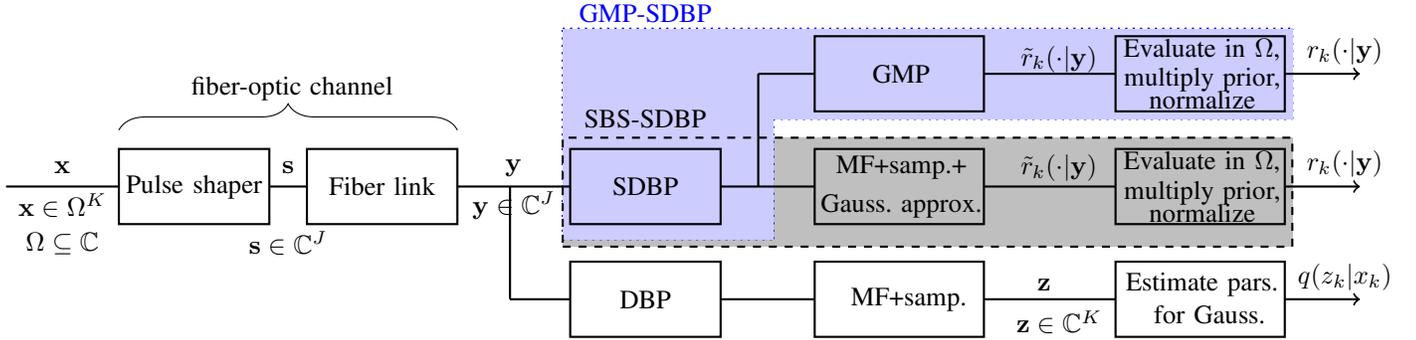


\section{Computation of AIR for the fiber-optic channel}\label{secAuxSDBP}

The computation of AIRs for the fiber-optic channel using different auxiliary channels is abstracted in  Fig.~\ref{figDiffAuxChanlSDBP}. The input data $\vecx$ is sent through a pulse shaper followed by the fiber link to get the output of the fiber-optic channel, $\vecy$. This output is fed either to SDBP or DBP, which aim to undo the impairments induced by the fiber-optic channel. The effect of pulse shaping is reversed using either the output from DBP or through one of the two techniques for SDBP to get the auxiliary channels. In the section, these processes will be detailed. 

\subsection{Computation of AIR using DBP}

The traditional approach of computing the AIR is by assuming the auxiliary forward channel to be memoryless\footnote{When the output of the channel $y_i$ at discrete time $i$, given the channel input $x_i$ at time $i$, is independent of channel inputs and outputs at all other times, we call the channel as memoryless.}. This assumption is justified by using a post-processing block, such as a DBP block for nonlinear compensation, after the fiber-optic channel as part of the auxiliary forward channel. The output statistics of the fiber-optic channel and the DBP has been considered memoryless with additive Gaussian noise. 
As shown in Fig.~\ref{figDiffAuxChanlSDBP}, let $\vecz=[z_1, z_2,\ldots,z_K]$, with $z_i \in \mathbb C$ for $i=1,2,\ldots,K$, be the signal after DBP, matched filtering and sampling. According to the {data-processing inequality} \cite[Ch.~2]{Cover2006}, the information content of a signal cannot be increased after post-processing and hence we have $I(\vecX; \vecZ)\leq I(\vecX; \vecY)$, where $$I(\vecX; \vecZ)=\mathbb E_{\vecX,\vecZ}\bigg[\log\frac{p(\vecZ|\vecX)}{p(\vecZ)}\bigg]$$ with $p(\vecz)=\int_{\vecx} p(\vecx)p(\vecz|\vecx)\mathrm{d}\vecx$. Similar to the lower bound (\ref{eqnIqyxleqIxy}) using auxiliary forward channel, we can define
\begin{align}\label{eqnIqzxleqIxy}
I_q(\vecX;\vecZ) \triangleq \mathbb E_{\vecX,\vecZ}\bigg[\log\frac{q(\vecZ|\vecX)}{q(\vecZ)}\bigg] \leq I(\vecX;\vecZ),
\end{align}
with $q(\vecz)=\int_{\alpX} p(\vecx)q(\vecz|\vecx)\mathrm{d}\vecx$.
Using DBP as detector, $q(\vecz|\vecx)$ is commonly assumed to factorize into marginal distributions, i.e., $q(\vecz|\vecx)=\prod_{k=1}^K q(z_k|x_k)$ and $q(\vecz)=\prod_{k=1}^K q(z_k)$.
By using these factorizations in (\ref{eqnIqzxleqIxy}) and using an equivalent of (\ref{eqnMIMC_qyx}) for $\vecz$ as output, we have
\begin{align}\label{eqnMCmemlessXZ}
\hat{I}_q(X;Z) = \frac{1}{\lenMC}\sum_{n=1}^{\lenMC} \left\{ \frac{1}{\lenX}\sum_{k=1}^{\lenX} \log\frac{q(z_k^{(n)}|x_k^{(n)})}{q(z_k^{(n)})} \right \},
\end{align}
where, similarly to the approach in \cite{Eriksson2016}, $q(z_k|x_k)$ is {assumed}\footnote{When superscripts for indicating MC run $n$ are omitted as in $q(z_k|x_k)$, it should be interpreted as applying to any general MC run.} to be {a Gaussian distribution with a different mean and covariance matrix for each possible value of $x_k$. In particular, the real and imaginary components of $z_k$ are taken to be either} independent and identically distributed Gaussian (iidG) {\cite{Secondini2013,Fehenberger2015a}}, or correlated Gaussian (CG) {\cite{Eriksson2016}. Gains in AIR were seen in the latter case for inline dispersion compensation at high powers. 
We will use both these approaches in benchmarking the results.}  In both these variations, a training phase is employed to obtain mean and variance corresponding to each of the constellation points. This training phase is visualized as the `Estimate pars. for Gauss.' block in Fig.~\ref{figDiffAuxChanlSDBP}, referring to the estimation of parameters for the Gaussian distribution. The means are obtained using \cite[Eq.~(8)]{Eriksson2016}, and variances for iidG and CG are obtained using \cite[Eq.~(9)]{Eriksson2016} and \cite[Eq.~(10)]{Eriksson2016}, respectively.

\subsubsection*{Remark 2}
{Note that even though AIRs for DBP are computed using an auxiliary forward channel, the same AIRs are also obtained using an auxiliary backward channel induced by this auxiliary forward channel, i.e., $r(\vecx|\vecz)= q(\vecz|\vecx)p(\vecx)/q(\vecz)$.}

\subsection{Computation of AIR using SDBP}

DBP is not an optimal processing strategy for nonlinear compensation. Indeed, some residual memory due to signal--noise interaction is present even after DBP is performed. As SDBP accounts for this memory, it may lead to {improved} bounds on the MI.

The theory behind SDBP is based on factor graphs and message passing, and is derived and explained in detail in \cite{Irukulapati2014TCOM}, while improved versions of SDBP are found in \cite{Irukulapati2014ECOC, Wymeersch2015SPAWC,Irukulapati2015JLT}. A short summary of SDBP is provided here for completeness.
SDBP compensates not only for linear and nonlinear effects existing in the fiber but also accounts for the noise from the amplifiers. The main idea of SDBP is to {statistically (i.e., in the form of a distribution) describe the uncertainty present in the unobserved signals at each stage of the fiber-optic channel. These unobserved signals are signals after each of the linear and nonlinear blocks of the split-step Fourier method (SSFM) and also the signals after the amplifiers. For the fiber-optic channel, since closed-form expressions of the distributions are not possible to derive, except for some specific scenarios, we represent distributions with a list of $\varNumPart$ particles.\footnote{{{A list of particles  $\mathbf{x}^{(1)},\ \mathbf{x}^{(2)},\,\ldots,\mathbf{x}^{(\varNumPart)}$, denoted by $\{\mathbf{x}^{(n)}\}_{n=1}^{\varNumPart}$, forms a particle representation of a  distribution $p(\mathbf{x})$ when $p(\mathbf{x})\approx 1/{\varNumPart} \sum_{n=1}^{\varNumPart}\delta(\mathbf{x}-\mathbf{x}^{(n)})$ \cite{Irukulapati2015JLT}. {We have taken uniform sampling of these particles to form a distribution. There exist other approaches such as nonuniform quantization \cite{Lin2015} to estimate the PDFs. }}}} 
 Each of these $\varNumPart$ particles is a waveform }{{ of size $J$ samples, where $J / K \approx 4 $, assuming waveforms are represented by $4$ samples per symbol. }
{Starting from the received signal $\vecy$, these $\varNumPart$ particles are passed through the inverse of each of the blocks of the fiber-optic channel, all the way to the transmitted signal $\vecs$.} {At each inverse amplification stage, a particle representation of the injected amplified spontaneous emission (ASE) noise is also required. This is obtained by drawing a random vector $\mathbf{w}$ comprising $N \times JN_{p}$ independent, identically distributed, and circularly symmetric complex Gaussian random variables, collecting the required $N_{p}$ particle representations, each of size  $J$, for each of the $N$  amplifiers.} 
SDBP can thus be viewed as an algorithm that takes an input $\vecy \in \mathbb C^{\lenY}$ and returns {$\varNumPart$ particles}, where each of these $\varNumPart$ {particles} is in $\mathbb C^{\lenY}$, and describes the knowledge the receiver has regarding the variable $\vecs$ in Fig.~\ref{figDiffAuxChanlSDBP}. {The algorithm is stochastic, as its result depends on the  realization $\mathbf{w}$ used to compute the particles. This dependence, whose impact can be made negligible by increasing $N_{p}$, will be omitted in the following. However, we will shortly come back to it at the end of this section, showing that it affects only the tightness of the computed lower bounds and not their validity.} To account for the effect of pulse shaping, two different approaches, SBS-SDBP and GMP-SDBP, are proposed in \cite{Irukulapati2014TCOM} and \cite{Wymeersch2015SPAWC}, respectively, and are explained briefly below.

In the first approach, SBS-SDBP, the output after SDBP is passed through a matched filter, {matched to the transmit pulse shape,} followed by sampling\footnote{There is residual memory left after SBS-SDBP \cite{Irukulapati2014ECOC} as matched filter followed by sampling is a linear technique and may not be the optimal processing for the nonlinear fiber-optic channel \cite{Liga2015,Agrell2015RSTA}. This residual memory was accounted for by using the Viterbi algorithm on the samples obtained after a matched filter, and was shown to have improved performance compared to SBS-SDBP \cite{Irukulapati2015JLT}.}. Corresponding to each symbol $x_k$, {$\varNumPart$ particles} are approximated with a multivariate Gaussian distribution, $\tilde{r}_k(\cdot|\vecy)$. In the second approach, referred to as GMP-SDBP in this paper, all {$\varNumPart$ particles} from SDBP are first approximated with a multivariate Gaussian distribution and then Gaussian message passing (GMP) is applied according to \cite[Table III]{Loeliger2007FG} instead of a matched filter, and $\tilde{r}_k(\cdot|\vecy)$ is obtained. 
In SBS-SDBP and GMP-SDBP, the distribution $\tilde{r}_k(\cdot|\vecy)$ is evaluated at $x_k \in \Omega$, {multiplied with the prior,} and normalized to get an auxiliary backward channel as shown in Fig.~\ref{figDiffAuxChanlSDBP}. By assuming that the input distribution $p(\vecx)$ {is the} product of its marginals, i.e., $p_{\vecX}(\vecx)=\prod_{k=1}^{\lenX}p_{X}(x_k)$ and by assuming that the auxiliary backward channel is factorized as
\begin{align}\label{eqnRxyRxky}
r(\vecx|\vecy) \triangleq \prod_{k=1}^{\lenX} r_k(x_k|\vecy),
\end{align}
(\ref{eqnMIMC_Rxy}) becomes
\begin{align}\label{eqnMIMC_Irxvecy}
\hat{I}_r^{\textrm{mem}} = \frac{1}{\lenMC}\sum_{n=1}^{\lenMC} \left\{ \frac{1}{\lenX}\sum_{k=1}^{\lenX} \log\frac{r_k(x_k^{(n)}|\vecy^{(n)})}{p_X(x_k^{(n)})} \right \},
\end{align}
where 
$r_k(x_k^{(n)}|\vecy^{(n)})$ is obtained by either SBS-SDBP or GMP-SDBP. 

\subsubsection*{Effect of Number of Particles in SDBP}
{
In order to make explicit the dependence of the auxiliary backward channel estimated by SDBP on $\mathbf{w}$, we indicate the auxiliary backward channel as $r(\cdot|\mathbf{y},\mathbf{w})$. Indeed, for a large enough number of particles $N_{p}$, we can assume that the particle representation of the ASE noise becomes very accurate and almost independent of $\mathbf{w}$. In this case, the auxiliary backward channel can be written as $r(\cdot|\mathbf{y},\mathbf{w})\approx r(\cdot|\mathbf{y})$ and directly used in (\ref{eqnIqxyleqIxy}) and (\ref{eqnMIMC_Rxy}). However, we might be interested in the practical case in which $N_{p}$ cannot be made large at will (e.g., due to complexity constraints), such that the auxiliary backward channel is affected by relevant statistical fluctuations and its dependence on $\mathbf{w}$ cannot be neglected. This case requires some extra care, as the lower bound (8) assumes that the auxiliary backward channel is fixed (though arbitrary). On the other hand, given the same output samples $\mathbf{y}$, the SDBP algorithm may provide different auxiliary backward channels, depending on the internally generated random vector $\mathbf{w}$. That is to say, when considering the MC average in the estimators (\ref{eqnMIMC_Rxy}) or (\ref{eqnMIMC_Irxvecy}), the detector is optimized for a different auxiliary backward channel at each MC run. Therefore, we generalize the lower bound (\ref{eqnIqxyleqIxy})  as 
\begin{equation}
I(\mathbf{X};\mathbf{Y})\ge I_{r}(\mathbf{X};\mathbf{Y}|\mathbf{W})\triangleq\mathbb{E}_{\mathbf{w}}[I_{r_{\mathbf{w}}}(\mathbf{X};\mathbf{Y}))]\label{eq:generalized_auxiliary backward channel_lower-bound}
\end{equation}
where $I_{r_{\mathbf{w}}}(\mathbf{X};\mathbf{Y})\triangleq I_{r}(\mathbf{X};\mathbf{Y}|\mathbf{W}=\mathbf{w})$
is the auxiliary backward channel lower bound (\ref{eqnIqxyleqIxy}) obtained for a fixed realization $\mathbf{w}$, while the quantity $I_{r}(\mathbf{X};\mathbf{Y}|\mathbf{W})$ can be interpreted as the average AIR when the mismatched detector is randomly selected (with some probability $p(\mathbf{w})$) from a family of detectors $r(\cdot|\mathbf{y},\mathbf{w})$. The inequality in (\ref{eq:generalized_auxiliary backward channel_lower-bound}) follows from the fact that
\begin{equation}
I_{r_{\mathbf{w}}}(\mathbf{X};\mathbf{Y})\le I(\mathbf{X};\mathbf{Y})
\end{equation}
for any $\mathbf{w}$ and holds regardless of the number of particles $N_{p}$. The average AIR in (\ref{eq:generalized_auxiliary backward channel_lower-bound}) can be eventually estimated as
\begin{equation}
\hat{I}_{r}^{\mathrm{mem}}=\frac{1}{N_{\mathrm{mc}}}\sum_{n=1}^{N_{\mathrm{mc}}}\left\{ \frac{1}{K}\log\frac{r(\mathbf{x}^{(n)}|\mathbf{y}^{(n)},\mathbf{w}^{(n)})}{p(\mathbf{x}^{(n)})}\right\} .\label{eq:generalized_MC_estimator}
\end{equation}
In conclusion, the number of particles $N_{p}$ does not affect the validity of the bound (\ref{eq:generalized_auxiliary backward channel_lower-bound}) and can be selected to obtain the desired trade-off between tightness
and complexity. For a small value of $N_{p}$, the statistical fluctuations of the auxiliary backward channel (from one MC run to the other) and its actual mismatch can be quite significant, but are anyway part of the adopted detection strategy. They are incorporated in the average AIR defined in (\ref{eq:generalized_auxiliary backward channel_lower-bound}) and are averaged out by the MC estimation (\ref{eq:generalized_MC_estimator}). As in (\ref{eqnMIMC_Irxvecy}), the computation of (\ref{eq:generalized_MC_estimator}) can be simplified by factorizing both the auxiliary backward channel and the input distribution.
}

\section{Numerical Results and Discussion}\label{secNumSim}

\subsection{System Parameters}

\begin{figure}
\centering
\psfrag{Tx}[c][c]{\scriptsize{Tx}}
\psfrag{SMF}[c][c]{\scriptsize{Single-mode fiber}}
\psfrag{EDFA1}[c][c]{\scriptsize{EDFA}}
\psfrag{DCM}[c][c]{\scriptsize{fiber Bragg grating}}
\psfrag{EDFA2}[c][c]{\scriptsize{EDFA}}
\psfrag{Rx}[c][c]{\scriptsize{Rx}}
\psfrag{Fiber-optical link}[c][c]{\scriptsize{Fiber-optical link}}
\psfrag{N}{{\scriptsize $\times \spanNum$}}
\psfrag{vt1}[c][c]{}
\psfrag{v1}[c][c]{}
\psfrag{rtf}[c][c]{}
\psfrag{r}[c][c]{}
\psfrag{Fiber-optical link}[c][c]{{Fiber-optical link}}
\psfrag{for DM link}[c][c]{\scriptsize{for DM link}}
\psfrag{a}{(a)}

\psfrag{pulseshaper}[bl][c][1][90]{\scriptsize{Pulse Shaper}}
\psfrag{SDBP}[bl][c][0.8][90]{\scriptsize{DBP/SDBP}}
\psfrag{Decsn}[bl][c][1][90]{\scriptsize{Decisions}}

\psfrag{M}{\scriptsize{x$M$}}
\psfrag{exp(jGU(t,z))}{\scriptsize $\exp{(j\gamma \Delta |.|^2)}$}
\psfrag{NonLinear}{\scriptsize{Nonlinear}}
\psfrag{Linear}{\scriptsize{Linear}}
\psfrag{HCD}{\scriptsize{$\text{H}_{\text{CD}}$}}
\psfrag{SMF2}[c][c]{\scriptsize{SMF/DCF}}
\psfrag{b}{(b)}

\psfrag{G}{\scriptsize$G_i$}
\psfrag{noise}{$\mathbf w_{ni}$}
\psfrag{EDFAj}{\scriptsize{EDFA1/EDFA2}}
\psfrag{c}{(c)}

\includegraphics[width=1\columnwidth]{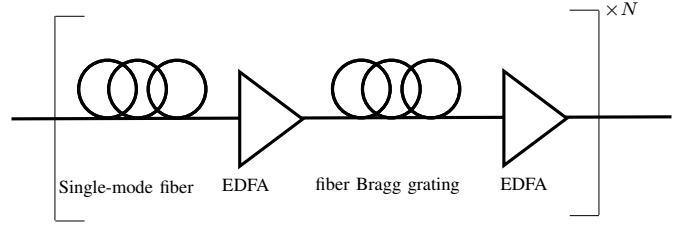}
\caption{\label{fig:FiberLink}\small{A fiber link with $N$ spans where each span consists of an single mode fiber, fiber Bragg grating, and erbium-doped fiber amplifers (EDFA).}}
\end{figure}

The fiber-optic channel used in this paper {is a single-channel system} comprising a single-polarization transmitter 
and a fiber link {consisting of $N$ spans. Each span of the fiber link consists of a transmission fiber of length $L$, which is a standard single-mode fiber simulated using SSFM, and a fiber Bragg grating for optical dispersion management. In between fiber spans, there are erbium-doped fiber amplifiers that compensate for the losses in the preceding span.} The transmitter uses a root raised cosine pulse shaper with a roll-off factor of 0.25 and truncation length of 16 symbol periods. The modulation format is 64-QAM and the symbol rate $R_\text{s}$ is either $14$ GBd, $28$ GBd, or $56$ GBd. The parameters used for the standard single-mode fiber are {a dispersion coefficient of $D=16$ ps/(nm km), a Kerr nonlinearity parameter of $\gamma=1.3~\text{(W km)}^{-1}$, and an attenuation of {$\alpha=0.2$ dB/km},} which are according to the ITU-T G.652 standard. Propagation in the fiber is simulated using the SSFM with a segment length \cite{Zhang2008stepSize} of $\Delta = (\epsilon L_\text{N} L_\text{D}^{2})^{1/3},$ where $\epsilon  = 10^{-4}$, $L_\text{N}=1/(\gamma P)$ is the nonlinear length, $L_\text{D} = {2 \pi c}/(R_\text{s}^{2}|D|\lambda^2)$ is the dispersion length, $\lambda$ is the wavelength, $c$ is the speed of light, and $P$ is the average input power to each fiber span. We used the same segment lengths for simulating the channel and for both DBP and SDBP. A fiber Bragg grating with an insertion loss of $3$ dB and perfect dispersion compensation for the preceding standard single-mode fiber is used.  The noise figure is {$F_\text{n}=5.5$} dB for each of the amplifiers. 
{The total noise power spectral density due to amplified spontaneous emission is $N_\text{ase}= ((G_1-1)n_\text{sp}h\nu+(G_2-1)n_\text{sp}h\nu)N$, where $G_1 (\text{resp. } G_2)$ is the gain in the amplifier after standard single-mode fiber (resp. fiber Bragg grating), $n_\text{sp}$ is the spontaneous-emission factor and is approximately $F_\text{n}/2$ when gains of the amplifiers are large, $h$ is Planck's constant, and $\nu$ is the optical frequency.} {Ideal band-pass filters with an equivalent low-pass bandwidth}
equal to the symbol rate are used in the erbium-doped fiber amplifiers and at the input  of the receiver.  The {number} of particles used in the SDBP approach is $\varNumPart=500$ for both SBS-SDBP and GMP-SDBP, {but we verified that even with $\varNumPart=1500$, similar performance was obtained.} {For simulations, the input distribution is assumed to be uniform, i.e., $p_X(x_k)=1/|\Omega|$ for $x_k \in \Omega$.}

\subsection{Results}

\begin{figure}%
    \centering
    {\includegraphics[width=.49\textwidth]
 {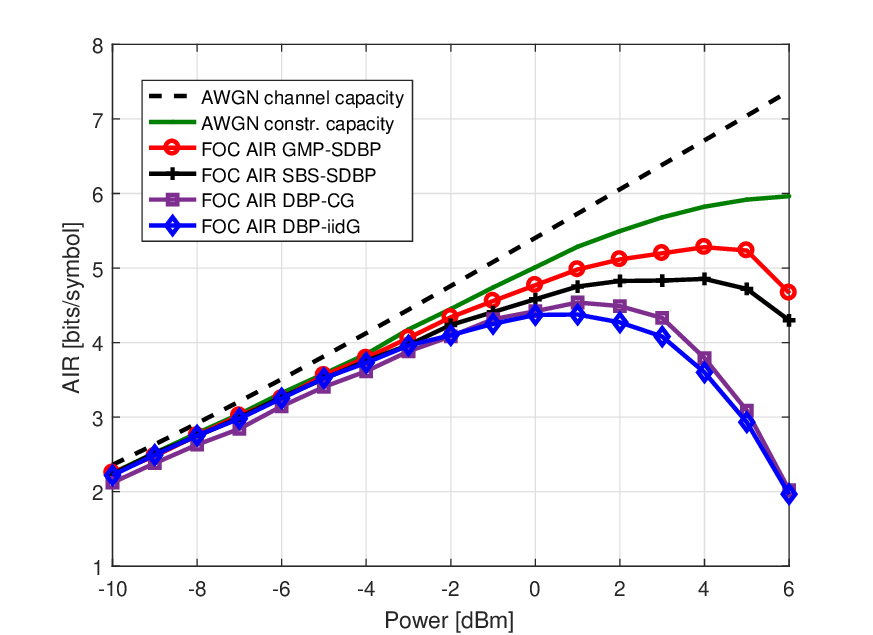}}%
    \caption{AIR using DBP-iidG {\cite{Secondini2013,Fehenberger2015a}}, DBP-CG {\cite{Eriksson2016}}, SBS-SDBP {\cite{Irukulapati2014TCOM}}, and GMP-SDBP {\cite{Wymeersch2015SPAWC}} for 14 GBd, 64-QAM, fiber Bragg grating link, $N=30$, $L=120$ km. {FOC refers to fiber-optic channel. }}
    \label{fig_AIRSDBP}%
\end{figure}

\begin{figure}%
    \centering
   {{\includegraphics[width=0.49\textwidth]
{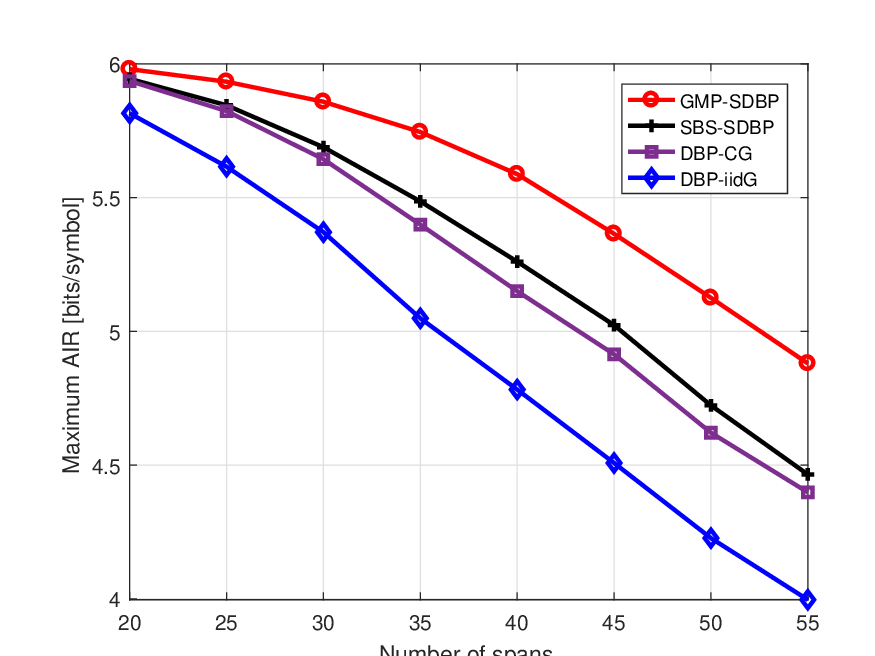} }}%
    \caption{{Maximum AIR (obtained at optimum power) for diferent number of spans for 14 GBd, 64-QAM, and $L=100$ km. }}
    \label{fig_AIRSDBP_2}%
\end{figure}

\begin{figure}%
    \centering
    {{\includegraphics[width=0.49\textwidth]
{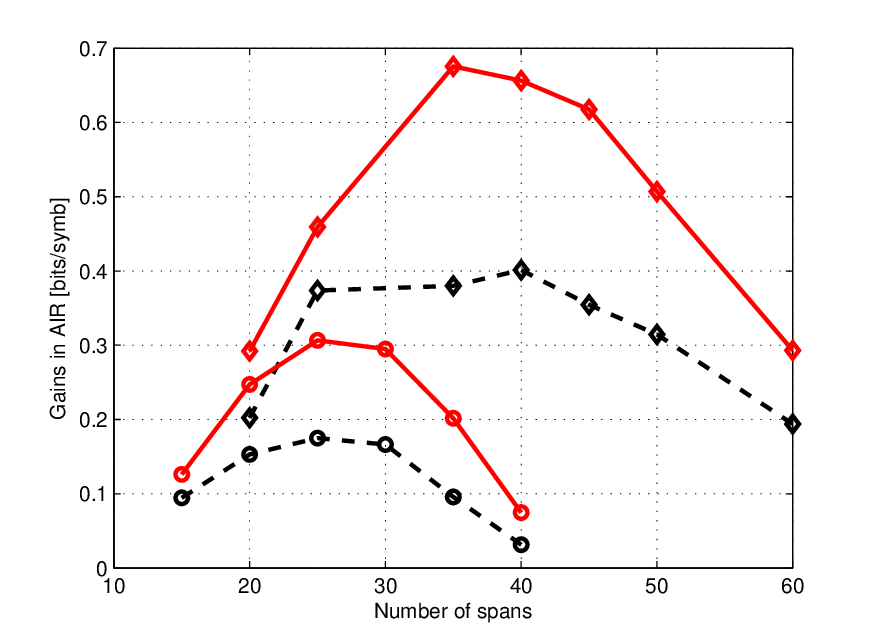} }}%
    \caption{{The gains in AIR of GMP-SDBP (resp. SBS-SDBP) over DBP-CG are shown using solid (resp. dashed) lines for 64-QAM, {$L=100$ km} for 28 GBd (diamonds)  and 56 GBd (circles). }}
    \label{fig_AIRSDBP_3}%
\end{figure}

Fig.~\ref{fig_AIRSDBP} shows the lower bounds on the MI, AIR, as a function of input power, obtained through different auxiliary channels for 14 GBd over a 64-QAM link. For reference, the capacity of the additive white Gaussian noise (AWGN) channel and the constrained capacity 
of an equivalent AWGN channel with 64-ary quadrature amplitude modulation (QAM) are also shown. We notice a trend for the DBP approaches (diamond and square markers) to have lower AIRs than the SDBP approaches (plus and circle markers), and also that the AIRs behave differently for DBP and SDBP. {Specifically, we can observe that the behavior of the link is almost linear up to about $-4$  dBm, beyond which the four AIR curves start to seperate. At higher powers, nonlinearity comes into play and the curves separate. SDBP reaches a maximum at around $4$ dBm, while DBP reachers a lower maximum at about $1$ dBm.}  The AIR for DBP decreases faster than that of SDBP, which means that SDBP performs better in the nonlinear regime. This is an expected behavior because SDBP accounts for the nonlinear signal--noise interactions, which DBP does not account for. If we compare AIRs between DBP techniques, DBP-CG (square markers) has better AIR than DBP-iidG (diamond markers), as the former accounts for the correlation between the in-phase and quadrature components. {This is expected, and is in line with the conclusions of \cite{Eriksson2016}, as DBP-CG has better AIR than DBP-iidG.} Also, we can observe that GMP-SDBP has better AIR than SBS-SDBP, as GMP-SDBP is a more principled way of computing a specific message, whereas SBS-SDBP is a heuristic approach. 

 {AIRs similar to Fig.~\ref{fig_AIRSDBP} were computed by varying the number of spans $N$  for 14 GBd, 64-QAM, and $L=100$ km. The maximum AIR for DBP-iidG, DBP-CG, SBS-SDBP, and GMP-SDBP is obtained at optimum power and plotted in Fig.~\ref{fig_AIRSDBP_2}}. 
It can be seen that as $N$ is increased, the AIR decreases. 
The highest AIR, which is the maximum that one can achieve for 64-QAM, is achieved at a low number of spans, which is $N=20$ for the chosen parameters. We can also observe that at $N=50$, the gain in AIR with GMP-SDBP over DBP-iidG (resp. DBP-CG) is 0.9 bits/symbol (resp. 0.7 bits/symbol) and the gain with SBS-SDBP is 0.5 bits/symbol (resp. 0.3 bits/symbol). } 

{AIRs similar to Fig.~\ref{fig_AIRSDBP_2} were computed for 28 GBd (diamond markers)  and 56 GBd (circle markers) for different $N$, and the maximum AIR is computed for these approaches. The difference between the maximum AIR for SBS-SDBP or GMP-SDBP and the maximum AIR for DBP-CG (which is used as a benchmark technique) is shown in Fig.~\ref{fig_AIRSDBP_3}.  All the curves have a maximum at some intermediate point as the gain between SDBP and DBP will be zero at small $N$ (both AIRs saturate to the maximum value), and at large $N$, both AIRs vanish. We observe that the gains in AIR of SDBP over DBP decrease as the symbol rate is increased. This behavior was observed also in our previous research \cite{Irukulapati2014TCOM, Irukulapati2015JLT}. The SDBP accounts for {signal--noise} interactions, which decrease as we increase the symbol rate. Therefore, the performance of SDBP approaches DBP with increasing symbol rate. As the symbol rate increases, accounting for longer correlations in the memory becomes even more important. However, due to complexity constraints, we have not taken  longer memory into account. 
With the recent interest in  symbol-rate optimization \cite{Poggiolini2015, Guiomar2017OFC}, low symbol rates are getting more attention, and therefore, gains obtained at 14 GBd for single-carrier transmission may still be relevant. } 

\subsection{Discussion}

The computation of lower bounds on the MI using either an auxiliary forward channel or an auxiliary backward channel are two different ways with their own advantages and disadvantages. If an auxiliary forward channel is available, then $q(\vecy)$ has to be calculated first, which may involve some integrals, and (\ref{eqnIqyxleqIxy}) is used to lower-bound the MI. Lower bounds on the MI can be obtained using an auxiliary backward channel by using any conditional distribution $r(\vecx|\vecy)$, i.e., by removing the constraint that $r_q(\vecx|\vecy)$ in (\ref{eqnKLDgeq0}) is induced by an auxiliary forward channel. That is, if an auxiliary backward channel is available, then (\ref{eqnIqxyleqIxy}) can be used to lower-bound the MI without explicitly finding an auxiliary forward channel. The true MI can in theory be obtained either by maximizing over all possible auxiliary forward channels in (\ref{eqnIqyxleqIxy}), or by maximizing over all possible auxiliary backward channels in (\ref{eqnIqxyleqIxy}). In this paper, we obtained auxiliary backward channels from the SDBP algorithm and the results indicate that {lower bounds on the MI computed using GMP-SDBP were the best in comparison to DBP-iidG, DBP-CG, and SBS-SDBP.}

We will discuss the extension of the results for dual polarization and comment on the complexity. Firstly, computation of AIRs through the auxiliary backward channel using (\ref{eqnMIMC_Irxvecy}) is applicable for dual polarization also. {SBS-SDBP has been developed for dual polarization \cite{Irukulapati2014TCOM} and GMP-SDBP can be extended to dual polarization. A single polarization was used for computing AIRs with GMP-SDBP in this paper for computational simplicity.} We note that the polarization mode dispersion for a dual-polarization transmitter degrades the performance of both DBP and SDBP, as observed in \cite{Irukulapati2014TCOM}, and hence the AIRs of Fig.~\ref{fig_AIRSDBP} will be lowered. However, we conjecture that the relative gains of SDBP compared to DBP would be similar to what we have shown in Fig.~\ref{fig_AIRSDBP} for a single polarization.
Secondly, the number of segments per span used in the simulation of the fiber using SSFM is the same as the ones used for DBP and SDBP. There exist many low-complexity variations of DBP, where the number of segments is optimized for real-time implementation. Low-complexity variations of SDBP can be derived, e.g., by optimizing number of particles or segments per span \cite{Irukulapati2014TCOM}. 

{Improved} lower bounds may possibly be obtained than those reported in the paper. Here we present two different methods. Firstly, in the GMP-SDBP, we used a linear Gaussian message passing algorithm to account for the effect of the pulse shaper, which may not be an optimal strategy for the fiber-optic channel. We conjecture that when the distribution is represented in a particle form, techniques other than linear Gaussian message passing might yield even {better} bounds than those presented in the paper. Secondly, {improved} bounds on the MI can be obtained by extending the principle used from auxiliary forward channel to auxiliary backward channel to a more general technique. The required {property} to lower-bound the MI using an auxiliary forward channel and also an auxiliary backward channel is $D\geq0$ in (\ref{eqnKLDgeq0}) and (\ref{eqnKLDRrgeq0}). This principle can be extended by allowing $p(\vecy)$ in (\ref{eqnKLDgeq0}) to be an arbitrary probability density function over $\vecy$ {that is not necessarily induced by $p(\vecy|\vecx)$ or $q(\vecy|\vecx)$} \cite{Sadeghi2009}. 

{
The methodology introduced in the paper for deriving the SDBP detector and computing the AIR applies to any fiber-optic system for which an SSFM-like channel description exists. Such systems may include, e.g., wavelength division multiplexing, spatial division multiplexing, arbitrary dispersion maps, arbitrary amplification schemes, and arbitrary modulation formats. For example, higher AIRs can be achieved with optimized signal constellations than with regular QAM \cite{Djordjevic2010}. Such system changes would affect all AIR bounds in a similar manner. } 

{\textit{Remark 3}: An upper bound for the mutual information is presented as $I(\vecX; \vecY) \leq L \log(1+\mathit{SNR})$ bit/block \cite[eq. (27)]{Kramer2015} where SNR is the signal-to-noise ratio. With the notation of this paper, $I(\vecX; \vecY) \leq (J/K) \log(1+(P/R_\text{s})/N_\text{ase})$ bit/symbol. Unfortunately, this upper bound scales with the oversampling factor $J/K$, whereas the lower bounds do not scale correspondingly. Already with the selected value of $J/K=4$, the upper bound is much larger than the computed AIRs and falls outside the ranges plotted in Figs. 3--5.}

\section{Conclusion}\label{secConcl}

Traditionally, lower bounds on the MI were computed using an auxiliary forward channel. In this paper, we computed lower bounds using an auxiliary backward channel for the first time for the fiber-optic channel. These bounds are achievable by a maximum a posteriori detector based on the auxiliary backward channel.
Two different distributions obtained through the SDBP algorithm are used as auxiliary backward channels for estimation of AIRs. Both these distributions have better AIR in comparison to the state-of-the-art method of using DBP. Through simulations, it was also found that up to {0.7} bit/symbol higher AIR is obtained using GMP-SDBP compared to DBP. This means that in comparison to the DBP approach, {improved} lower bounds on the MI can be obtained using the SDBP approach. 

\section{Acknowledgments}
The authors would like to thank Rahul Devassy, Kamran Keykhosravi, Giulio Colavolpe, Tobias Fehenberger, Amina Piemontese, and Alex Alvarado for fruitful discussions. {The authors would also like to thank anonymous reviewers for their comments, which improved the presentation of the paper. }

\bibliographystyle{IEEEtran}
\bibliography{C:/Naga/Dropbox/library}

\newpage

\end{document}